\documentclass[12pt]{JHEP3}
\usepackage{amsmath,epsfig}
\def\be{\begin{equation}}
\def\ee{\end{equation}}
\def\ba{\begin{eqnarray}}
\def\ea{\end{eqnarray}}

\def\beq{\begin{equation}}
\def\eeq{\end{equation}}
\def\beqa{\begin{eqnarray}}
\def\eeqa{\end{eqnarray}}

\def\preal{{\rm Re\,}}
\def\pim{{\rm Im\,}}

\def\yzero{\smash{\hbox{$y\kern-4pt\raise1pt\hbox{${}^\circ$}$}}}
\def\p{\partial}
\def\a{\alpha}
\def\b{\beta}

\def\beq{\begin{equation}}
\def\eeq{\end{equation}}
\def\beqa{\begin{eqnarray}}
\def\eeqa{\end{eqnarray}}

\def\-{\hphantom{-}}

\def\s2{\frac{1}{\sqrt2}}

\def\oh{\frac{1}{2}}
\def\beq{\begin{equation}}
\def\eeq{\end{equation}}
\def\beqa{\begin{eqnarray}}
\def\eeqa{\end{eqnarray}}

\def\Tr{{\rm Tr \,}}

\def\IF{\relax{\rm I\kern-.18em F}}
\def\II{\relax{\rm I\kern-.18em I}}
\def\IP{\relax{\rm I\kern-.18em P}}
\def\IC{\relax\hbox{\kern.25em$\inbar\kern-.3em{\rm C}$}}
\def\IR{\relax{\rm I\kern-.18em R}}

\def\Dsl{\,\raise.15ex\hbox{/}\mkern-13.5mu D} 
\def\IZ{Z\kern-.4em  Z}

\def\fn{\footnote}

\def\inte{{\bf Z}}

\def\cA{{\cal A}}

\def\ti{\tilde}

\def\pd{{\rm PD}}

\newcommand{\drawsquare}[2]{\hbox{%
\rule{#2pt}{#1pt}\hskip-#2pt
\rule{#1pt}{#2pt}\hskip-#1pt
\rule[#1pt]{#1pt}{#2pt}}\rule[#1pt]{#2pt}{#2pt}\hskip-#2pt
\rule{#2pt}{#1pt}}

\newcommand{\fund}{\raisebox{-.5pt}{\drawsquare{6.5}{0.4}}}
\newcommand{\Ysymm}{\raisebox{-.5pt}{\drawsquare{6.5}{0.4}}\hskip-0.4pt%
        \raisebox{-.5pt}{\drawsquare{6.5}{0.4}}}
\newcommand{\Yasymm}{\raisebox{-3.5pt}{\drawsquare{6.5}{0.4}}\hskip-6.9pt%
        \raisebox{3pt}{\drawsquare{6.5}{0.4}}}
\newcommand{\antifund}{\overline{\fund}}

\title{Moduli Stabilisation and de Sitter String Vacua from Magnetised D7
Branes}

\author{D. Cremades$^1$, M.-P. Garc\'{\i}a del Moral$^2$, F. Quevedo$^1$,
K. Suruliz$^1$ \\
$^1$ DAMTP/CMS, University of Cambridge,\\ Wilberforce Road, Cambridge CB3
0WA, UK.\\ \\
$^2$ Dipartimento di Fisica Teorica, Universit\'a di Torino and
I.N.F.N, Sezione di Torino, Torino,\\  via P. Giuria 1, I-10125
Torino, Italy} \abstract{Anomalous $U(1)$'s are ubiquitous in 4D
chiral string
  models. Their presence crucially
affects the process of moduli stabilisation and cannot be neglected in
  realistic set-ups. Their net effect in the 4D effective action is to
induce a matter field dependence in the non-perturbative
  superpotential
and a Fayet-Iliopoulos D-term. We study flux compactifications of IIB
  string theory in the  presence of magnetised D7 branes. These give
  rise to
   anomalous $U(1)$'s that modify the standard moduli stabilisation
procedure. We consider simple orientifold models to determine the
  matter field spectrum and the form of the effective field theory.
We apply our results to  one-modulus KKLT and multi-moduli
  large volume scenarios, in particular to the Calabi-Yau
  $\IP^4_{[1,1,1,6,9]}$.
 After stabilising the matter fields, the
  effective action for the
  K\"ahler moduli can acquire an extra positive term that can be used for
  de Sitter lifting with non-vanishing F- and D-terms. This provides  an
explicit realization of the D-term lifting proposal of \cite{hepth0309187}.
}

\preprint{DAMTP-2006-22\\DFTT-02/2007\\ hep-th/0701154}

\begin{document}

\section{Introduction}
Significant progress has been made recently regarding the
supersymmetry breaking and moduli stabilisation problems of string
compactifications (for recent reviews see \cite{hepth0610102,hepth0610327}).
In particular, IIB string flux compactifications have provided
concrete models of moduli stabilisation in which the scale of
supersymmetry breaking can be calculated as well as the relevant soft
breaking terms in the effective action.

This has been achieved independently of the details and location of the
chiral fields of the Standard Model. This is in part because of the
`modular' structure inherent in type II  string models. There are
global bulk issues such as moduli stabilisation, inflation and supersymmetry
breaking,
and there are local brane issues regarding the gauge group,
spectrum of chiral fields, etc. Usually the two types of issues
can be approached independently of each other. Separating global questions
from local ones in a systematic construction of realistic models was
proposed in \cite{aiqu} and called the `bottom-up' approach
to string model building. For relevant progress on the model building
side see \cite{cgqu,hepth0508089}.

This procedure is very efficient in the sense that once a global
problem, such as moduli stabilisation, has been solved, then
a realistic D-brane construction in terms of D-branes at singularities
or magnetised D7 branes can be attached to the
compactification manifold to make it into a realistic model.
However at the end of the day we have to consider the two parts
together in order to control the low-energy nature of soft supersymmetry
breaking
terms, reheating, local and global symmetries, etc. We therefore
should investigate
the effects
that chiral D-brane models have in the effective action for the moduli
fields and if it is needed to incorporate them in the moduli stabilisation
procedure.

One of the generic properties of chiral D-brane models is the presence
of anomalous $U(1)$'s. In a typical construction, the chiral matter of
the spectrum naturally induces anomalies for some of the $U(1)$ gauge
fields. The anomaly is cancelled by the standard Green-Schwarz
mechanism with the net effects of giving a mass to the corresponding
gauge field, a `charge' to the
modulus field corresponding to the gauge coupling of the effective
field theory, and inducing a Fayet-Iliopoulos D-term proportional to the
total charge of the chiral fields.  This shows in particular that
the non-perturbative terms in the KKLT scenario \cite{ hepth0301240} involving only
the K\" ahler modulus
field are not gauge invariant, and therefore that the chiral matter fields
must also enter the superpotential in such a way to render it gauge invariant.
Thus the effects of anomalous $U(1)$ fields must be taken into account in the
moduli stabilisation procedure if we have chiral fields living in
D-branes, as is required for instance by the inclusion of the Standard Model.

The fact that anomalous $U(1)$'s also induce Fayet-Iliopoulos (FI)  D-terms
can modify the moduli stabilisation procedure. Since D-terms are
positive, it was proposed in \cite{hepth0309187} that they can be used
to lift the original KKLT AdS minimum to a de Sitter one in a way
consistent with a supersymmetric effective action. See
\cite{hepth0503216,fabio,beatriz,hepth0607077} for
recent discussions of this proposal.
More generally there is an important question that emerges here: could
the effects induced by anomalous $U(1)$'s change the successful results
regarding moduli stabilisation so far?

In this article we address the issue of moduli stabilisation in the
presence of anomalous $U(1)$'s\footnote{For a perturbative moduli
  stabilisation proposal including FI terms see \cite{mp}.}. Following the standard procedure (and
the proposal of \cite{hepth0309187}) we consider D7 branes with
non-vanishing magnetic fluxes. Magnetic fluxes are the standard source
of chirality and of the anomalous $U(1)$'s. In the next section
we consider simple
orientifold models and determine the corresponding $U(1)$ charges of
the different matter fields and  compute the FI term. In section 3 we
 consider simple examples
of one modulus KKLT type or several moduli large volume type and see
how moduli stabilisation is affected by these new ingredients,
including a
non-perturbative superpotential of the Affleck-Dine-Seiberg type
invariant under the anomalous $U(1)$. We find that these effects
either leave the good features of the model, such as the existence of
exponentially large volumes, or modify them to allow de
Sitter lifting, depending on the model, the distribution of the D7
branes and magnetic fluxes. We include several appendices with
details of some of the calculations, including anomaly cancellation and the FI
term.

\section{D-terms and de Sitter vacua}
\subsection{General considerations}
One of the key steps in the KKLT procedure is the lifting of vacuum
energy to a positive value.
In the original KKLT paper \cite{hepth0301240}, the lifting term
arose from an anti-D3 brane localised in a highly warped region of
the Calabi-Yau. This type of brane breaks supersymmetry explicitly,
and the low energy effective theory cannot be described by the standard $N=1$
four dimensional supergravity.

An alternative and more controlled lifting mechanism was
proposed in \cite{hepth0309187}.
The idea is to use a D-term generated by magnetic fluxes on D7 branes
to provide an additional positive contribution
to the scalar potential. This can, under favourable conditions, result in
a de Sitter vacuum. The formalism is that of supergravity and the supersymmetry
breaking is spontaneous rather than explicit.

One of the reasons the original proposal of \cite{hepth0309187} ought to be
studied in more detail is the observation, pointed out in \cite{hepth0503216},
that
in a general $N=1$ supergravity theory, there exists a relation
between F-terms and D-terms,
\begin{equation}
\label{dtermfterm}
D = {i\over W} \sum_i (\delta \phi_i) D_i W.
\end{equation}
Here $W$ is the superpotential, $f$ the gauge kinetic function,
$D_iW=\partial_i W +W\partial_iK$
the K\"ahler covariant derivative with $K$ the K\"ahler potential
and  $\delta \phi_i$ the transformation of the field $\phi_i$ under
the $U(1)$ generating the D-term.
 The relation (\ref{dtermfterm}) follows simply from gauge invariance
and holds at any point in field space (except where $W=0$). It is clear that
$D_i W = 0$ implies $D=0$, making it essentially impossible to uplift
an original SUSY vacuum, i.e.
to have pure D-term supersymmetry breaking, contrary to the standard
global supersymmetry case.
The proposal in   \cite{hepth0309187} actually considered both F and D
terms to be non-vanishing once matter fields were introduced, but an
explicit analysis was not done. This has been done  in recent
publications \cite{fabio,beatriz,hepth0607077} in the context of the
KKLT scenario.

Since the original KKLT scenario is such that the AdS minimum is
supersymmetric, it is difficult to simply lift it by a standard D-term.
Usually $D=0$ can be used to fix the charged matter fields $\phi_i$
giving the KKLT F-term potential as a function of the moduli fields only,
and hence essentially recovering the KKLT result.

On the other hand there exist a large class of models in
which moduli stabilisation is achieved in
such a way that the AdS minimum is non-supersymmetric
\cite{hepth0408054, hepth0502058,hepth0505076}. In most of these cases
the volume is exponentially large with very interesting
phenomenological implications. Since the
original F-term is non-zero  it is natural to expect that
a D-term can  be non-zero, contributing a positive term to
the scalar potential and therefore leading to the possibility of de
Sitter lifting.
Therefore
the natural context in which to look for vacua that can be uplifted with
D-terms is that of nonsupersymmetric large volume compactifications.
Another interesting question for these models is whether the addition of
D-terms still preserves the fact that the stabilised volume is still
exponentially large.

D-terms in string theory \cite{dsw}
have a very interesting structure, since they are generically related to
massive and anomalous $U(1)$s and the Green-Schwarz mechanism, and this
relation
greatly constraints the form of the D-term potential.
We will briefly review this topic in the following subsection, before
considering concrete setups in which we will carry out the lifting procedure.

\subsection{On FI terms, anomalous U(1)s and massive gauge bosons}

As explained in \cite{qsusy,hepth0502059,hepth0609211}, turning on internal
magnetic flux in the worldvolume of a D7 brane
generates an FI term in the four dimensional world-volume theory. This can be
seen from several points of view. Firstly, the presence of the
magnetic field generates chiral fermions and scalars in the low energy theory.
For certain values of the moduli the scalars become massless and the theory
supersymmetric. This pattern of supersymmetry breaking/restoration is clearly
reminiscent of an FI mechanism.
But perhaps the clearest way to see that an FI term
is generated is considering the fact that four dimensional couplings of the
form $\int D_2\wedge F$, where the 2-form $D_2$ comes from the reduction of the
4-form RR field $C_4$, and $F$ is the four dimensional gauge field strength, are
generated when internal magnetic flux is turned on.
This kind of four dimensional coupling generates masses for the
corresponding $U(1)$ gauge bosons. The 2-form field $D_2$ is the four
dimensional Hodge dual of an axion zero form $\phi$, which is the imaginary
part of a
chiral superfield modulus $T$ that parametrises the volume of some 4-cycle. The
field $\phi$ transforms under a $U(1)$ gauge transformation as
$\phi\to\phi+Q\theta(x)$, where $Q$ (the charge of this transformation) is
related to the internal magnetic flux on the D-brane.
It follows that in order to
get the $\int D_2\wedge F$ coupling, the K\"ahler potential
of the four dimensional theory must depend on the combination
$T+T^*+QV$, with $V$ the vector multiplet corresponding to $F$ \cite{dsw}. But
in a supersymmetric theory the presence of a gauge boson mass is always linked
to an FI term, since both come from the same term in the Lagrangian when
expressed in the superfield formulation. For global supersymmetry:
\beqa
\int d^4\theta~K(T+T^*+QV)=\left({\p K\over\p V}\right)_{V=0}V|_{\theta^4}+\oh
\left({\p^2K\over \p V^2}\right)_{V=0}(\p_\mu\phi+A_\mu)^2+...\nonumber
\eeqa

Let us be more precise. Consider the Chern-Simons part of the D-brane action
\be
\label{d7cs}
S_{CS} =-\mu_7\int_{D7} \sum_p \imath^* C_p \wedge e^{\imath^* B + {\cal{F}}}.
\ee
Here ${\cal{F}}$ is a mass dimension two form, and ${\cal{F}}\equiv 2\pi
\alpha' F$ so that $F$ has mass dimension 0.
$\imath^*$ denotes the pullback operation.
Expanding this action, the coupling to the RR 4-form $C_4$ is\fn{We will not
consider couplings to the NSNS field $B_2$ in what follows.}
\begin{equation}
\int_{D7} C_4 \wedge F\wedge F.
\end{equation}
Taking one of the $F$'s to be the compact flux $f$
and the other to be with non-compact indices (denoted by $F$), and
reducing $C_4 = D_2^\alpha \wedge \omega_\alpha+\cdots$, where $\omega_\alpha$
form a basis
for the 2-cohomology of the Calabi-Yau, we have
\begin{equation}
\label{coupling}
\int_{\Sigma} \omega_\alpha\wedge {f} \int_{M^4} D_2^\alpha \wedge F.
\end{equation}
Here $\Sigma$ denotes the 4-cycle wrapped by the D-brane. What is interesting
to note here is that since $D_2^\alpha$ is
correlated with the 2-cycle whose volume form is proportional to
$\omega_\alpha$, its four dimensional Hodge dual, the axion zero form
$\sigma_\a=\int_{\Sigma_\alpha}C_4$, must be related
to the four cycle that is Poincar\'e dual of $\omega_\alpha$, namely
$\ti\omega^\a$. Since one defines the K\"ahler moduli fields as \cite{hepth0609211}
\begin{equation}
T_\alpha = {1\over{2\pi(2\pi \sqrt{\alpha'})^4}g_s} \left( \int_{\Sigma_\alpha}
\sqrt{g} d^4 x + i \sigma_\alpha \right),
\end{equation}
we see that the K\"ahler moduli fields that get charged under the $U(1)$ are
those parameterising volumes of a four-cycles that have non-zero intersection
with
the two-cycle where the magnetic field is supported. Also, since the coupling
(\ref{coupling}) gives a mass for the $U(1)$ gauge boson supported on the brane
with internal magnetic flux, it follows from supersymmetry that whenever
magnetic flux is turned on in a D7-brane, both a Fayet-Iliopoulos
term\fn{Note that since this FI term is a field dependent quantity, it can
indeed be
zero in some regions of the moduli space. If this was not the case, it would be
impossible to have magnetised brane
constructions preserving supersymmetry.} and a
mass for the corresponding $U(1)$ gauge boson are generated. However, it is
important to emphasise that this magnetic flux does not generically
generate a charge for the $T$ modulus whose vev parametrises the volume of the
4-cycle wrapped by the D7 with internal magnetic flux.
This field $T$ only gets charged if the corresponding 4-cycle has
self-intersections \cite{hepth0502059,hepth0609211}.

The fact that a given field $T_\a$ gets charged does not mean that the $U(1)_a$
associated to the brane supporting magnetic flux is anomalous.
Such a $U(1)$ will
be anomalous if there exists a term in the four dimensional low energy
effective action of the form $\sigma_\a \Tr F_b\wedge F_b$,
where $F_b$ is the field strength of some $U(1)_b$ or $SU(N_b)$ gauge group
present in the construction, or
$\sigma_\a \Tr R\wedge R,$ $R$ being the Ricci form. The former couplings give
rise to mixed $U(1)_a-SU(N_b)^2$ and mixed $U(1)_a-U(1)_b^2$ anomalies
in the low energy theory (note that the cubic $a=b$ case is a particular case
of this). The latter couplings give rise to
gravitational anomalies in the low energy theory. There will be mixed gauge
anomalies for a given $U(1)$ whenever any D7 brane wraps a four-cycle that is
charged under this $U(1)$. The cubic anomalies will arise as a particular case
of this, either because the cycle has self-intersections or has a zero
intersection with its orientifold image. The gravitational anomaly will
typically arise when the corresponding D7 brane has non-zero intersection
with the orientifold. Note that every anomalous $U(1)$ will automatically be
massive, but a massive $U(1)$ need not be anomalous.

Consider a setup in which magnetic flux has been turned on in the
world-volume of a D7 brane wrapping a 4-cycle with K\" ahler modulus $T_F.$
The D-term potential arising from this setup is
\begin{equation}
V_D = {g^2\over{2}} \left( {Q_\alpha \over{4\pi^2}} \partial_{T_\alpha} K +
\sum_iq_i \phi_i K_{\phi_i}\right)^2.
\end{equation}
where the $T_\alpha$ are all the K\" ahler moduli charged under the
anomalous $U(1)$, the $\phi_i$ are the unnormalised open string fields charged
under the same $U(1)$, and $K_{\phi}=\p K/\p\phi$. One can get the expression
for the gauge coupling constant $g^2$ from the dimensional reduction of the DBI
action. The result
\cite{hepth0609211} is
\be
\label{gkf}
g^{-2} = {\rm Re} T_F - f_{F} {\rm Re} S.
\ee
Here $f_{F}$ is a certain magnetic flux dependent factor.
In the large $T_F$ limit which we will be concerned with throughout the
paper, we may neglect the ${\rm Re} S$ contribution to $g^{-2}.$



\begin{figure}[ht]
\linespread{0.2}
\begin{center}
\epsfxsize=0.60\hsize \epsfbox{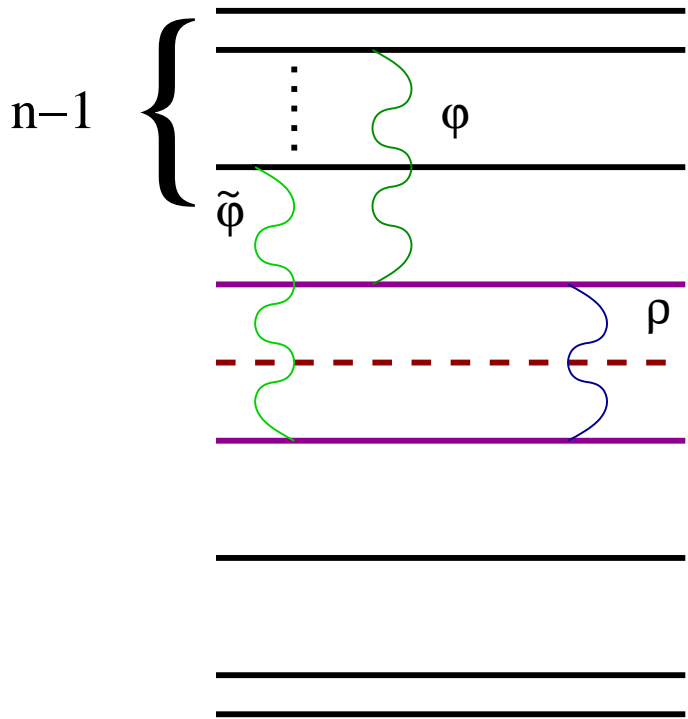}
\end{center}
\caption{ This is a representation of the basic set-up. A set of $n$ D
branes is such that one of them is magnetised leading to a
$U(n-1)\times U(1)$ gauge group. The dotted lines represent the
orientifold plane. Chiral fields are $\varphi$ in the fundamental of
$SU(N_c)$ (with $N_c=n-1$) corresponding to strings going from the
non-abelian set of branes to the magnetised brane, $\tilde{\varphi}$ in the
anti-fundamental,
representing strings with endpoints in the non-abelian set of branes
and the orientifold image of the magnetised brane and $\rho$
corresponding to strings with endpoints at the magnetised brane and
its orientifold image.  }
\label{setup}
\end{figure}

\subsection{General setup and spectrum}
\label{spectrumsec}
One of the  aims of this paper is to study under which circumstances it is
possible to get D-term lifting in string models. As already emphasised, it is
better to start
from a non-supersymmetric vacuum in order to get the lifting. Given this, a
natural context to work in
are the large volume models \cite{hepth0502058,hepth0505076}, in which
vacua have
naturally non vanishing F-terms.
One could also attempt to apply D-term lifting on KKLT vacua, since there may
still
be minima in which both the D-term and the F-terms are non-vanishing.
We can therefore imagine a few different scenarios. In the next
sections we will consider
the following:
\begin{enumerate}
{\item One may consider the simplest model with one K\" ahler modulus $T$,
with a nonperturbative superpotential fixing the modulus, and turn on
magnetic fluxes such that $T$ becomes charged.\label{poss1}}
{\item Alternatively, one can consider a model with at least two K\" ahler
moduli
for which large volume minima have been found \cite{hepth0502058}.
Then one may turn on magnetic fluxes on the exponentially large 4-cycle which
determines the volume so that
the corresponding modulus becomes
charged, without the need of including a nonperturbative superpotential for
it.\label{poss2}}
{\item One can also consider a model with more than one modulus and turn on
magnetic
fluxes on branes wrapping one of the `small' 4-cycles which is
nonperturbatively stabilised.
There are two sub-cases here, the first being that the cycle becoming charged
under the $U(1)$ is the large 4-cycle, and the second that the small 4-cycle
becomes charged. \label{poss3}}
\end{enumerate}

There is one further choice that must be made -- whether the D7 branes lie on
top of orientifold
planes or away from them. The former choice ensures that the local dilaton
charge is cancelled
but introduces new complications due to extra matter fields being present. We
consider
both possibilities.

In each one of these constructions, the local setup we will consider is as
follows. We will take a set of D-branes and O-planes, typically on
top of each other in order to cancel local tadpoles\fn{One can also consider
the branes being away from the orientifold plane but,
as emphasised in \cite{hepth0609211}, this potentially leads to F-theory
corrections that are beyond the scope of this work.}. We consider magnetic flux
turned
on in one of the branes in the stack. The presence of this magnetic flux will
generate an FI term in the four dimensional theory, as explained in the
previous
section. Also, the magnetic flux will be responsible for the appearance of
chiral superfields in the overlapping region with the magnetised two-cycle.

The gauge group will be of the form $G\times U(1)_F$, where the $U(1)_F$ factor
corresponds to the branes where we have put magnetic flux\fn{For some
orientifold
projections, the Chan-Paton projection will require to put the same units of
magnetic flux in more than a stack of branes in order to get the $U(1)$
factor.} (we
will call it brane or stack $F$)
and $G$ can be either\fn{This will depend on the orientifold projection over
the Chan-Paton
degrees of freedom. Generically, if the stack of branes has no magnetic flux
one will get
$SO(N)$ or $USp(N)$, with $N$ related to the number of branes in the stack in a
model-dependent way,
and the choice of gauge group depends on the choice of orientifold projection.
The $U(N)$ case can arise either if one considers the branes to be away from
the
orientifold projection, or if the stack of branes is also magnetised, with an
internal magnetic flux that is different from the flux of the D-brane
giving rise to the $U(1)_F$ gauge group.} $SO(N)$, $USp(N)$ or $U(N)$, and
lives in the remaining stack of branes (stack $G$).
In this situation, one expects the following set of fields to appear:
\begin{itemize}
\item{A set of fields transforming in the $(+1_F,\antifund_G)$ under
$U(1)_F\times G$, that live
between the brane $F$ and the stack $G$. We will call these fields
$\varphi_i$.}
\item{Another set of fields transforming in the $(+1_F,\fund_G)$ under
$U(1)_F\times G$, that live
between the magnetised brane $F$ and the orientifold image of $G$. We will
denote these fields by\fn{Note that there can be
situations in which $\fund_G=\antifund_G$ and hence the $\tilde \varphi$ fields
can be considered as the antiparticles of the $\varphi$, so that we do not need
to consider them in the D-term potential.} $\tilde \varphi_i$.}
\item{A set of fields $\rho$ with charge $\pm 2$ under $U(1)_F$. These fields
live between the brane $F$ and its orientifold image $F^*$.}
\end{itemize}

More generically, one can consider a stack of $F$-branes, all of them with the
same internal magnetic field. In this case, the fields $\varphi$ transform under
the
$(\fund_F,\antifund_G)$, the fields $\ti \varphi$ (if present) transform under
the $(\fund_F,\fund_G)$ and the $\rho_{ij}$ transform under symmetric and
antisymmetric
representations of $U(N_F)$ (note $N_F$ is not necessarily equal to the number
of $F$-branes). As we will see now, one can constrain the number of these
fields
from anomaly cancellation arguments.

\subsection{Anomaly cancellation constraints on the spectrum}
The setup we are considering is a set of D7-branes on top of a set of O-planes,
and we will turn on
magnetic flux on the former. This magnetic flux induces a tadpole for a lower
dimensional brane charge, that induces a chiral
anomaly in the world volume of the branes. Also, gravitational anomalies will generically be present.
We will compute the spectrum
on these branes using anomaly cancellation
arguments.

Consider two D7 branes $a$ and $b$, wrapping different 4-cycles $\Sigma_a$ and
$\Sigma_b$,
and overlapping over a 2-cycle where both of them support magnetic flux. This
gives rise to a number
of chiral fermions, given by the index
\beqa
I_{ab}&=&\int_{\Sigma_a} {\rm PD}_{Y}(\Sigma_b)\wedge
F_a-\int_{\Sigma_b}\pd_Y(\Sigma_a)\wedge F_b\nonumber\\
&=&\int_{\Sigma_b} {\rm PD}_{Y}(\Sigma_a)\wedge
F_a-\int_{\Sigma_a}\pd_Y(\Sigma_b)\wedge F_b
\label{inter}
\eeqa
where we define $\pd_A(B)$ to be the Poincar\'e dual of the cycle $B$ taken
with respect to the
(sub)manifold $A$.
In order to make contact with standard anomaly inflow arguments, we would like
to define this index as a product between D7 and D5 charges, such that the
product is
bilinear in them. To do this we define the total class of a D7 with magnetic
flux as
\cite{hepth0508089}
\beqa
[D7_a]=\pd_Y(\Sigma_a)+\pd_Y\left(\pd_{\Sigma_a}(F_a)\right).
\label{charges}
\eeqa
This definition includes the D7 charge specified by the (co)homology class
of the 4-cycle $\Sigma_a$ wrapped by the D7, and the D5 charges generated by
the world-volume magnetic
flux\fn{There is also a D3 charge that we do not mention since is irrelevant in
the following.}.
Then we define the inner product between two D7 classes as
\beqa
[D7_a]\cdot[D7_b]&\equiv&\int_Y
\left[\pd_Y(\Sigma_a)+\pd_Y\left(\pd_{\Sigma_a}(F_a)\right)\right]\wedge\left[\pd_Y(\Sigma_b)-\pd_Y\left(\pd_{\Sigma_b}(F_b)\right)\right]\nonumber\\
&=&\int_{\Sigma_b}\pd_Y\left(\pd_{\Sigma_a}(F_a)\right)-\int_{\Sigma_a}\pd_Y\left(\pd_{\Sigma_b}(F_b)\right).
\label{prod}
\eeqa
This intuitive definition is bilinear in the brane charges and applies in the
simplest cases.
In the general case several subtleties arise and one can define such a product
in a
much more formal (and complicated) way \cite{hepth0508089,douglas,aspinwall}.
Since we do not want to overload the paper with mathematical jargon, and
moreover since the only
property of $I_{ab}$ that will be relevant in what follows is bi-linearity in
the D5 and D7 charges (that is also present in the complete formulae),
we will use the simplified expression
(\ref{prod}) and refer the interested reader looking for a more formal
definition of the index to the above-mentioned papers.

As already mentioned, the total number of chiral multiplets charged under a
given gauge group will be given by the multiplets $\varphi$, $\ti\varphi$
appearing between the brane (or stack of branes)
supporting the gauge group and other (stacks of)
branes appearing in the configuration (and their orientifold images), plus a
set of charge $\pm 2$ fields (that we denote generically by $\rho$) that can
appear between the brane (or stack of branes) and its orientifold image. The
number of $\varphi$, $\ti\varphi$ fields in the configurations will be given by
the expressions
(\ref{inter}) or (\ref{prod}), while the number of $\rho$ fields is generically
model dependent. However, as we will see now, standard anomaly cancellation
arguments already used
in \cite{csu} can easily give us this number.

To start with, let us consider a configuration with a $U(N_F)\times U(N_G)$
gauge group. Generically we will have $N_F=1$ and $G\neq U(N_G)$, but since
what we want to
compute is the number of $\rho$ fields, whose number should only depend on the
relation between the brane $F$ and the orientifold and thus should be
independent of $G$, and moreover
most of the orthogonal and symplectic groups do not give rise to anomalies,
this calculation will suffice to compute the actual number of $\rho$ fields
regardless of the
particular situation.

Given the spectrum, the kind of anomalies one can expect are: $SU(N_F)^3$,
$U(1)_F-SU(N_G)^2$, $U(1)_F-SU(N_F)^2$, $U(1)_F-U(1)_G^2$,
$U(1)_F^3$ and
gravitational. Out of all these, only the first two must be cancelled without
any help from a Green-Schwarz mechanism since there are no $U(1)$s involved. In
this section we will
see how the matter content required by the cancellation by the cubic $SU(N_F)^3$ anomalies uniquely fixes the
number (and charge) of the
$\rho$-fields. We show in an appendix how this matter content is precisely the
one required for the rest of the anomalies to be cancelled, together
with a Green-Schwarz mechanism.

To get a $U(N_F)\times U(N_G)$ gauge group one has to put different magnetic
flux in two different stacks of branes.
Let us assume that we have a set of $N=n+m$ branes,
so that we put a flux of the form $F_n$ in the $n$ branes (that wrap a 4-cycle
$\Sigma_n$) and flux of the form $F_m$ units in the $m$ branes (that wrap a
four cycle $\Sigma_m$, not
necessarily different from $\Sigma_n$). More concretely, using the notation
defined in (\ref{charges})
\beqa
[D7_i]=\pd_Y(\Sigma_i)+\pd_Y\left(\pd_{\Sigma_i}(F_i)\right).
\eeqa
with $i=n,m$.
Tadpole cancellation implies
\beqa
n([D7_n]+[D7_n'])+m([D7_m]+[D7_m'])-[O7]=0.
\label{tadpoles}
\eeqa
with $[O7]$ the class of the orientifold plane, defined analogously to
(\ref{charges}) and where we have included the number of orientifold planes and
charge of the orientifold
compared to that of a D7 in the definition of $[O7]$.  $[D7_i']$ the
orientifold image of $[D7_i]$, given by
\beqa
[D7_i']=\pd_Y(\Sigma_i)-\pd_Y\left(\pd_{\Sigma_i}(F_i)\right).
\eeqa
The gauge group on the branes will be $U(N_F)\times U(N_G)$, with
$n/N_F=m/N_G=b\in \inte^+$. Now, we know that tadpole cancellation conditions
must imply the cancellation of cubic
$SU(k)$ anomalies for $k=n,m$, without any contribution from GS terms.

Let us consider the spectrum. There is a set of $\varphi$ fields transforming
in $(\fund_{N_F},\antifund_{N_G})$,
and a set of $\ti \varphi$ fields transforming in $(\fund_{N_F},\fund_{N_G})$.
The number of them is given by\fn{The absolute
value of this index signals the net number of fermions and the sign denotes its
four dimensional chirality. Since in field theory one typically choses all
fermions to be of the same chirality, one takes the convention in which a
positive sign for $I_{ab}$
indicates the existence of $|I_{ab}|$ fermions with positive chirality and
charges $(1_a,-1_b)$, whereas a negative sign for $I_{ab}$ implies the
existence of
$|I_{ab}|$ fermions with positive chirality and charges $(-1_a,1_b)$.}
\beqa
\# \varphi &\equiv& I_{nm} = [D7_n]\cdot[D7_m]\\
\# \ti \varphi &\equiv& I_{nm'} = [D7_n]\cdot[D7_m'].
\eeqa
There can also be symmetric and antisymmetric representations. The number of
these fields has to be related to the numbers
$I_{kk'}$ and $I_{kO}$, with $k=n,m$, defined as
\beqa
I_{kk'}&\equiv&[D7_k]\cdot[D7'_k],\\
I_{kO}&\equiv&[D7_k]\cdot[O7].
\eeqa
Thus, we express the number of symmetric and antisymmetric representations of
$N_F$ as
\beqa
\#\Ysymm_{N_F}&=&\a_sI_{nn'}+\b_sI_{nO}\label{nsym}\\
\#\Yasymm_{N_F}&=&\a_aI_{nn'}+\b_aI_{nO}.\label{nasym}
\eeqa
A similar formula holds for $N_G$. Multiplying eq. (\ref{tadpoles}) by $[D7_n]$
on the left we get
\beqa
nI_{nn'}+m(I_{nm}+I_{nm'})-I_{nO}=0.
\eeqa
Since $m(I_{nm}+I_{nm'})$ is precisely $b$ times the anomaly produced by the
$\varphi$, $\ti \varphi$, it follows that
\beqa
b\left[\cA_s (\#\Ysymm)+\cA_a (\#\Yasymm)\right]=nI_{nn'}-I_{nO}.
\label{anom}
\eeqa
with $\cA_{s(a)}$ is the anomaly produced by the (anti)symmetrics. Given that
$\cA_s=N_F+4$, $\cA_a=N_F-4$, substituting (\ref{nsym}) and (\ref{nasym}) into
(\ref{anom}), and requiring the
number of fields to be independent of $N_F$, we get
\beqa
\# \Ysymm_{N_F}&=&\oh(I_{nn'}-{1\over 4b}I_{nO}),\label{rho1}\\
\# \Yasymm_{N_F}&=&\oh(I_{nn'}+{1\over 4b}I_{nO})\label{rho2}.
\eeqa
These (anti)symmetrics are, from the $U(1)_{N_F}$ point of view, a set of
charge $\pm2$ fields
that we will call $\rho$.

There are $N_F(N_F +1)/2 \times\# \Ysymm_{N_F}$ of them coming from the
symmetric representation and
$N_F(N_F -1)/2\times\#\Yasymm_{N_F}$ coming from the antisymmetric one. The
total number of $\rho$ fields is then given by
\beqa
\#\rho={N_F^2\over 2} I_{nn'}-{N_F\over 8b}I_{nO}
\label{nrho}
\eeqa
Note that, depending on the orientifold content, this number can be positive,
negative or zero. In the following we will take the convention that the four
dimensional chirality
of the corresponding fermion is fixed and the sign of the number of $\rho$s is
equal to the charge of the field. Finally, we insist in the fact that since the
number of $\rho$
fields should be independent of the gauge group $G$, we will also have the
number of $\rho$ fields given by (\ref{nrho}) in the general situation with
$SO(N)$ or $USp(N)$ gauge
groups.

Summarising, we have found that there are in general three types of
fields, fundamentals $\varphi_i, \tilde{\varphi_j}$ and singlet fields under
the (generically non-abelian) group $G$,
$\rho_{ij}$, where $i,j$ are flavour indices.

Given this, one can always choose to fix the charges both of the $\varphi$ and $\ti\varphi$ to be $+1$.
Depending on the model under consideration, the field $\rho$ can have charge $+2$ or $-2$ (cf (\ref{nrho})). 
Also, anomaly cancellation arguments imply that the charge of $T$ has the opposite sign to the charge of both $\varphi$ and $\ti\varphi$. Given this,
the relative sign between the FI term and the $\varphi$, $\ti\varphi$ dependent parts of the D-term is uniquely determined by the sign of $\p_TK$.
Each one of these possibilities will lead to different behaviour regarding the D-terms
and potential de Sitter lifting. We will consider all these cases
in the following section.

\section{Explicit Constructions}

We now study how D-term lifting can be explicitly realised in the various
cases listed in the previous section.

\subsection{One modulus case}
The simplest case to consider is the one K\"ahler modulus case,
studied in the original
KKLT paper \cite{hepth0301240}. We will discuss it first for
illustration purposes.

We will assume that the 4-cycle corresponding to the only K\" ahler modulus $T$
carries D7 branes with a gauge theory undergoing gaugino condensation.
We will also assume that the stack of branes under consideration does not
intersect any other stacks of D3 branes, to avoid the appearance of
additional massless matter in the low energy theory (this possibility
has been recently discussed in \cite{baumann}).

We consider one of the D7s and turn on a $U(1)$ magnetic flux on a 2-cycle
which belongs to the 2-homology of the 4-cycle the branes wrap.
The fields $\rho$ corresponding to strings stretching between the magnetised D7
and its orientifold image have charge $\pm 2$ under the
$U(1)$ gauge group \fn{In general $\rho$ is a $N_f\times N_f$ matrix in flavour
space but
for simplicity we will first  restrict to a single field $\rho$, valid
strictly for $N_f=1$ but capturing the main physics for the general
case (we will discuss later the possible implications of the other
fields $\rho$ in the $N_f\neq 1$ cases)}. There are
also quark fields $\varphi, \tilde{\varphi}$ with charge $+1$ which correspond
to open strings
stretched between the magnetised brane and the unmagnetised D7s and
their orientifold images.

If an appropriate topological condition is satisfied (that the cycle wrapped by
the D7 branes
intersects itself over the 2-cycle with magnetic flux), the modulus $T$ will
become charged under the $U(1)$
and a D-term potential is induced of the form
\begin{equation}
\label{dtermpot}
V_D = {1\over{T+T^*}} \left( (\partial_{T} K) \delta T +
(\partial_\rho K) \delta \rho + (\partial_\varphi K) \delta \varphi \right)^2.
\end{equation}

If the $\rho$ fields have charge $+2$ then $D$ is positive definite
and cannot vanish. This clearly provides an extra lifting term of
order $(T+T^*)^{-3}$ to the scalar potential precisely as proposed in
\cite{hepth0309187}. This option was recently studied in
a supergravity model motivated by string theory in \cite{beatriz}.

Let us now consider what is probably the most generic case of a
 $\rho$ field with charge $-2$. In this case the $\varphi$ fields are
massive and can be integrated out and we are only left with a tachyonic field
$\rho.$ This can be done since the ${\varphi}$ fields have charge $+1$ (same
sign as the FI term). Also there is a mass term in the superpotential
of the form $\rho \ {\varphi \varphi}$ which gives a mass to ${\varphi}$ once
the field $\rho$
of charge $q=-2$ gets a vev. If the mass of ${\varphi}$ is greater than the
effective value of the renormalisation group invariant scale $\Lambda$
of the non-abelian gauge theory, the fields ${\varphi}$ can be integrated out
and we are left with an effective theory in terms of only $\rho$ and
the K\"ahler modulus $T$.

We start then with the 4D supersymmetric effective action in terms of
one matter field $\rho$  with charge $q=-2$ and one K\"ahler modulus
$T$ with an anomaly induced charge $Q/(4\pi^2).$ Before eliminating the
$\varphi$ fields, the Affleck-Dine-Seiberg superpotential has the form
\be
W_{np} = \left( {\Lambda^{3N_c-N_f}\over{\det
(\varphi\tilde{\varphi})}}\right)^{1\over {N_c-N_f}}
\ee
where $\Lambda^{3N_c-N_f} = e^{-8\pi^2 T}$ is the scale of
gaugino condensation,
$N_c$ is the number of colours determined by the number of un-magnetised D7
branes and $N_f$ the number of flavours of chiral fields $\varphi$.
This is gauge invariant if $Q=-N_f.$

After integrating out the $\varphi$ fields, the Lagrangian is
determined by the following superpotential and K\"ahler potential\fn{$W_0$ is the standard flux superpotential. $\xi$ is the effect of the $\alpha'$ corrections as computed in 
\cite{bbhl}, that is proportional to the Euler number of the Calabi-Yau.}

\begin{eqnarray}
W=W_{0}+\rho^{a}e^{-bT}\\
K=-2\log (\tau^{3/2}+\xi)+\frac{\vert \rho\vert^{2}}{\tau^{2/3}},
\label{crap}
\end{eqnarray}
where $W_0$ is the constant flux superpotential, $a,b$ are constants
determined by the requirement that $W$ has dimension three and is
gauge invariant. So:
\begin{equation}
a=\frac{N_f}{N_c} \qquad b=\frac{8\pi^2}{N_c}.
\end{equation}
We have also defined $2\tau=T+T^*$ and added the $\alpha'$ correction to
the K\"ahler potential specified by $\xi$. In (\ref{crap}) we have used the
recent result in \cite{CCQPaper} regarding the moduli dependence of
the matter field K\"ahler potential indicating a `modular weight' of
$-2/3$ for the matter field $\rho$.

The scalar potential as a function of $T$ and $\rho$ is of the form:
\be V=V_F+V_D \ee with \be V_F=
e^{K}(D_{i}W\overline{D_{j}W}K^{i\overline{j}}-3|W|^{2}), \qquad
V_{D}=\frac{g^2}{2} D^2=\frac{1}{2\tau} \left(\frac{3Q}{8\pi^2
\tau}-\frac{q\vert\rho\vert^{2}}{\tau^{2/3}}\right)^{2}\ \ee by
neglecting both the matter contribution to $\partial_T K$ in $D$, which goes as
$-\frac{Q\vert\rho\vert^{2}}{12\pi^2\tau^{5/3}}$, and the $\alpha'$ correction parametrised by $\xi$.
In principle we need to find the extrema of this potential for the
two complex (four real) fields $T,\rho$. For the axionic parts
$\theta_T={\rm Im}T$ and
$\theta_\rho= {\rm Arg} \rho$, we can see that they only appear in the
scalar potential through the non-perturbative part of $W$ and therefore
only in the combination $\Theta=a\theta_\rho - b\theta_T$. The orthogonal
combination does not appear in the potential at all but it is just as well,
since this is precisely the combination that is eaten by the anomalous
$U(1)$ gauge field to get a mass as it can be easily verified.
Extremising with respect to $\Theta$ is straightforward (since it only
appears in the scalar potential through $\cos\Theta$, with extrema
at $\Theta=m\pi$).
Therefore the relevant fields to concentrate on are $\tau$ and the
modulus of $\rho.$

Since
\begin{eqnarray}
D_{\rho}W&=&a\rho^{a-1}e^{-bT}+\rho^{*}\tau^{-2/3}W\\
D_{T}W&=&-b\rho^{a}e^{-bT}-\left[\frac{3}{2(\tau+\xi
\tau^{-1/2})}+\frac{\vert\rho\vert^{2}}{3\tau^{5/3}}\right]W
\end{eqnarray}
we can easily see that
\be
q\rho \frac{D_\rho W}{W} + {Q\over{4\pi^2}} \frac{D_TW}{W}= D.
\ee
Therefore, for a KKLT like scenario in which $W_0$ is very small and there
are solutions to $D_TW=D_\rho W=0$, it can be seen that $D=0$ automatically.
In practise it is easier to solve for $\rho$ in  $D=0$ and substitute
in $D_TW=0$ to solve for $\tau$. This immediately reduces this system to
the KKLT one with no relevant effect from the D-term. This is as
expected from the discussions by Choi et {\it al} \cite{hepth0503216}.
We thus conclude that it is safe to ignore the effects of anomalous
$U(1)$'s in the KKLT scenario, where the original AdS vacuum is
supersymmetric. In this case both F and D terms vanish and we still
have a supersymmetric AdS vacuum.


We will  now consider a scenario similar to that of \cite{hepth0408054}, with a
one K\" ahler
modulus Calabi-Yau manifold, and turn on RR and NS fluxes such that the
flux-induced
superpotential is $W_0\sim 1$ which is more generic. Therefore the
nonperturbative
effect stabilising the K\" ahler modulus $T$ is much smaller in
magnitude than $W_0.$
Then the supersymmetry preserving condition $D_TW=0$ cannot be
satisfied and there is no KKLT minimum.
In a  large $\tau$ approximation, assuming that $W_0$ dominates over the
nonperturbative terms in $V_F$, and after minimising the phase of
$\rho$ and the imaginary part of $T$, the expression for $V_F$
becomes:
\be
V_F \sim
\frac{\vert
W_{0}\vert^{2}\vert\rho\vert^{2}}{3\tau^{11/3}}\left[1+\frac{2\vert\rho\vert^{2}}{3\tau^{2/3}}\right]
\ +\ \frac{\xi \vert
W_0\vert^{2}}{\tau^{9/2}}\ +\ V_{nonpert}
\ee
 Where $V_{nonpert}$ is the non-perturbative part of $V_F$.

In order to minimise the full scalar potential with respect to $\rho$,
we observe that for large $\tau$ the D-term dominates
and the minimum takes the form:
\be
\vert\rho\vert^{2}=\frac{3Q}{8\pi^2 q\tau^{1/3}}(1+\epsilon),
\ee
with $\epsilon\sim \frac{1}{\tau}.$
Substituting this result in $V_{D}$ gives
\be
V_{D}=\frac{9Q^{2}\epsilon^{2}}{2(8\pi^2)^2 q^{2}\tau^{3}}\sim O(\tau^{-5}),
\ee
and $V_{F}$ behaves as
\be
\label{este}
V_{F}\sim\vert W_{0}\vert^{2}\frac{Q}{8\pi^2 q\tau^{4}}  +\ \frac{\xi \vert
W_0\vert^{2}}{\tau^{9/2}}\ +\ V_{nonpert}
\ee
Where \be V_{nonpert}\ =\ Ae^{-2b\tau} - BW_0e^{-b\tau} \ee and $A$
and $B$ are functions of inverse powers of $\tau$. At leading order
they are
\begin{eqnarray} A&=&\frac{b^2 Q^a }{q^a
(\frac{\tau}{2})^{1+a/3}}+\frac{Q^{a}(6b-4/3ab+3a^2\frac{q}{Q})}{q^{a}(\frac{\tau}{2})^{2+a/3}}\nonumber
  \\
B&=&\frac{6bQ^{a/2}}{q^{a/2}(\frac{\tau}{2})^{2+a/6}}+\frac{aQ^{a/2}}{q^{a/2}(\frac{\tau}{2})^{3+a/6}}. \end{eqnarray} 
Notice that the
second and third term of expression (\ref{este}) are as in the
standard KKLT scenarios with added $\alpha'$ corrections
\cite{hepth0408054}. The net effect of the $\rho$ field is adding
the first term (which dominates over the D-term at large $\tau$).
This term is precisely what we need in order to lift to de Sitter
space. If without including this term the minimum for $\tau$ is an
AdS one, the potential at that minimum scales as $-1/\tau^{9/2}$.
The first term in (\ref{este}) dominates over this at large $\tau$
since it scales as $+1/\tau^4$. Furthermore, since both powers are
similar we need only a tuning of order $1/\tau^{1/2}$ in the
coefficient of the $\rho$ induced $V_F$ in order to have a de Sitter
minimum (and not wash away completely the original AdS minimum). The
de Sitter minimum can be obtained for values of $\tau\sim 10-100$.
The necessary tuning is smaller than in the original KKLT case due
to the higher power of the lifting term. Still the magnitude of this
term can be controlled if the 2-cycle where the magnetic flux is  at
the tip of a warped throat as in the KKLT case. Notice that in this
case the parameter $Q$ would be modified by the warp factor as
expected \cite{hepth0309187}.

This is therefore an explicit realisation of de Sitter lifting from D-terms
as proposed in \cite{hepth0309187}. Notice that both D and F terms are
non-vanishing in the resulting minimum. Notice however that in the original
discussion of \cite{hepth0408054} there were also de Sitter minima, even though
at relatively small volume. The D-term lifting does not appear to be
particularly useful in this case. Things are different in the more
generic, many K\"ahler moduli case.

\subsection{Two moduli case}
Let us consider the toy model of a hyper-surface in the weighted projective
space $\IP^4_{[1,1,1,6,9]}$ studied in \cite{hepth0505076,hepth0404257},
with two K\" ahler moduli $T_b$ and $T_s.$

The K\"ahler potential for the K\"ahler moduli is given by
\be
K=-2\log\left({\cal{V}}+\xi\right)
\qquad
{\rm with}\qquad
{\cal{V}}=\left(T_b+T_b^*\right)^{3/2}-\left(T_s+T_s^*\right)^{3/2}.
\ee

We assume that a stack of D7 branes wraps the small
4-cycle, such that a nonperturbative superpotential is generated in the
four dimensional effective field theory. This will generically result in
stabilising the modulus $T_b$ corresponding to the overall volume
perturbatively at an exponentially large value while $T_s$ will be
fixed at an ${\cal{O}} (1)$ value. For details of the construction
we refer to \cite{hepth0502058}.

\subsubsection{Fluxes on the large cycle}
Let us turn on magnetic flux on a 2-cycle
which is a sub-cycle of the large 4-cycle corresponding to $T_b.$

For simplicity we will assume that the modulus $T_b$ is entirely perturbatively
stabilised, so that it does not appear in the superpotential (this is
justified a posteriori if the volume is exponentially large we can
neglect the nonperturbative dependence on $T_b$).
This will be the case as long as there are a sufficient number of massless
adjoint
multiplets living on the world-volume of the branes left after turning
on RR and NS fluxes.
This means that the superpotential can be written as
\be
\label{superpot}
W = W_0 + A_s e^{-a_s T_s}.
\ee
We are assuming that $T_s$ is not charged under the $U(1)$ living on branes
wrapping the large cycle (which will be the case if the 4-cycles corresponding
to $T_s$ and $T_b$ do not intersect over any 2-cycle), so that we do not
need to include open string fields in the superpotential formula
$(\ref{superpot}).$

As explained in \cite{hepth0502058},
the nonperturbative effects  do not destabilise the
flux-stabilised complex structure and dilaton moduli. Therefore we
will only discuss the K\"ahler moduli dependence of the scalar potential.
For the time being, we assume an arbitrary parametrisation for the
K\" ahler potential of the $\rho$ field:
\begin{equation}
\label{kansatz}
K = -2 \log ({\cal{V}}) + c {{\rho \rho^*}\over{(T_b+T_b^*)^\alpha}}.
\end{equation}
It is easy to see that under a redefinition of $\rho$, $\rho\to
\rho/\sqrt{c}$, the dependence on $c$ is eliminated both from the
K\" ahler potential and from the D-term. Therefore we may set $c=1.$

The metric on moduli space is computed in appendix \ref{metricsappendix}. It
can be used to compute the full scalar potential, in particular
the lowest order F-term contribution involving $|\rho|^2$ which will
be of interest to us.

Let us investigate where $\rho$ is stabilised. If there were no
F-term contributions to its potential, $\rho$ would be stabilised by
the requirement that the D-term vanish. However, there are F-term contributions
coming from $K^{\rho\bar{\rho}} (D_\rho W) \overline{(D_\rho W)}$, as well
as $K^{b\bar{b}} (D_b W) \overline{(D_b W)}$ and the mixed terms
$K^{i\bar{\rho}} (D_i W) \overline{(D_\rho W)}$ (and complex conjugates).
All these contributions turn out to scale as the same power of $1/{\cal{V}}$:
\begin{equation}
{1\over{\cal{V}}^{2+{2\over3}\alpha}} |\rho|^2 |W_0|^2.
\end{equation}

The coefficient of this term can be shown to be $(1-\alpha).$ This is
done in Appendix \ref{coefcomp}. 
As discussed earlier, the sign of the charge of the $\varphi$ fields
is aligned with the sign of the FI term, so they always have a positive
mass, while the charges of $\varphi$ and $\rho$ can be either of the same or
opposite sign.
Let us first assume that the sign of the charge of the $\varphi$'s is the
same as the charge of $\rho.$

Therefore the D-term now has the form
\begin{equation}
V_D = {1\over{T_b+T_b^*}} \left( {p\over{T_b+T_b^*}} +
k \sum_i {|\rho_i|^2\over{(T_b+T_b^*)^\alpha}}
+ l \sum_j {|\varphi_j|^2\over{{(T_b+T_b^*)^\alpha}}} \right)^2,
\end{equation}
There are also F-term contributions to the scalar potential involving
$\rho_i$ and $\varphi_j$, of the form
\begin{equation}
\sum_i { |\rho_i|^2\over (T_b+T_b^*)^{3+\alpha}} + \sum_j {|\varphi_j|^2\over
(T_b+T_b^*)^{3+\alpha}}.
\end{equation}
The full potential is clearly minimised for $\varphi=\rho=0$, which gives an
uplift potential of the form
\be
V_D = {p^2\over {(T_b+T_b^*)^3}} \sim {p^2\over{{\cal{V}}^2}}.
\ee
This will be sufficient to uplift a nonsupersymmetric minimum with cosmological
constant $\sim -1/{\cal{V}}^3$ as long as there is a fine tuning in
$\epsilon=p^2$ of
order $1/{\cal{V}}.$ Recalling that $p = Q/(4\pi^2)$, with $Q$ an integer, we
can see
that $p^2$ is naturally of order $1/1000$, so for volumes in the region
${\cal{V}}\sim 10^3$
no extra fine tuning (coming from warping or other mechanisms) is required.
Note that this is a realisation of the original proposal of
\cite{hepth0309187}.

In this analysis we did not include the non-abelian D-term for the
quark fields,
\begin{equation}
V_D^{{\rm nonab}} = {1\over{{\rm Re} f}} \sum_a \left(
(\partial_{\varphi_i} K) T_{ij}^a \varphi_j \right)^2,
\end{equation}
with $T^a$ the generators of the nonabelian gauge group.
$V_D^{{\rm nonab}}$ is a positive definite contribution to the energy
which only depends on the $\varphi$ fields, so $\varphi=0$ is still a minimum.

Suppose now that the charge of the $\rho$ fields is opposite in sign to the
charge of
the $\varphi$ fields. Then the scalar potential is minimised at $\varphi=0$ and
only one of
the $\rho$ fields (say, $\rho\equiv \rho_1$) is nonzero. That there is a
stationary point with
these properties is clear. That it is a minimum can also easily be shown, after
observing that, after minimising,
\begin{equation}
D = {p\over{T_b+T_b^*}} - k {|\rho|^2\over{(T_b+T_b^*)^\alpha}} + l
{|\varphi|^2\over{(T_b+T_b^*)^\alpha}} =
{1\over{2k (T_b+T_b^*)^2}}.
\end{equation}
Then one has
\begin{equation}
{\partial^2 V\over{\partial \varphi \partial \varphi^*}} =
{1\over{(T_b+T_b^*)^{3+\alpha}}} +
{2l\over{(T_b+T_b^*)^{\alpha+1}}} D > 0.
\end{equation}
The $\rho_j, j>1$ are flat directions of $V.$

The D-term contribution to the mass of $\rho$ arises from
\begin{equation}
{1\over{T_b+T_b^*}} \left( {r\over (T_b + T_b^*)}  -
 { {q |\rho|^2 \over{(T_b+T_b)^{\alpha}}}}  \right)^2,
\end{equation}
and scales as $|\rho|^2/{\cal{V}}^{4/3+2\alpha/3}.$ Here any ${\cal{O}} (1)$
constant factors from differentiating $K$ have been absorbed into $q,r>0$ and
the negative sign has been explicitly inserted to emphasise that $\rho$
is tachyonic.

Thus the minimum for $\rho$ will be such that the D-term is almost,
but not completely, cancelled. More explicitly, let us consider
the $|\rho|^2$ contributions from the F- and D-terms:
\begin{equation}
V = f(T_s, T_b) +
{(1-\alpha)\over{\cal{V}}^{2+{2\over3}\alpha}} |\rho|^2 |W_0|^2 +
{1\over{T_b+T_b^*}} \left( {r\over (T_b + T_b^*)}  -
 { q {|\rho|^2 \over{(T_b+T_b)^{\alpha}}}} \right)^2.
\end{equation}
Minimising with respect to $\rho$ we find
\begin{equation}
{r\over (T_b + T_b^*)}  - {{q |\rho|^2 \over{(T_b+T_b)^{\alpha}}}}
 = {(1-\alpha)\over{2q}}{1\over{\cal{V}}^{4\over3}} |W_0|^2,
\end{equation}
so that
\begin{equation}
| \rho|^2 \sim {\cal{V}}^{2(\alpha-1)/3},
\end{equation}
and the F-term contribution to the scalar potential,
${(1-\alpha)\over{\cal{V}}^{2+{2\over3}\alpha}} |\rho|^2 |W_0|^2$, scales as
${\cal{V}}^{-8/3}$, while the D-term contribution scales as
${\cal{V}}^{-10/3}.$

Let us now investigate the resulting scalar potential, setting $\alpha=2/3$ as explained in section 3.1., and considering
only lowest order terms in the $1/{\cal{V}}$ expansion\fn{The constants $\mu$, $\nu$ and $\xi$ are defined as follows $\mu \sim a_s |A_s W_0|$, $\lambda \sim a_s^2 |A_s|^2$.
$\nu\sim |W_0|^2$, $\xi \sim -\chi(M)|W_0|^2$, with $\chi(M)$ the Euler number of the Calabi-Yau. See \cite{hepth0502058} for details.}

\begin{equation}
\label{potential}
V = {\lambda \sqrt{\tau_s} e^{-2 a_s \tau_s}\over {\cal{V}}} -
{\mu \over {\cal{V}}^2} \tau_s e^{-a_s \tau_s} + {\nu\over {\cal{V}}^{8/3}}
+ {\xi \over{\cal{V}}^3}.
\end{equation}
It turns out that neglecting the leading order
$\alpha'$-correction, which scales as $\xi/{\cal{V}}^3$,
the potential has no minima. There is a stationary point
determined by
\begin{eqnarray}
e^{a_s \tau_s} &=& \left( {3^6\over{8^3\cdot 2^5 }} \right)
\left( {{a_s A_s\over{W_0}}} \right) \tau_s^4\\
{\cal{V}} &=&  {3^6\over 2^{15}} \sqrt{\tau_s} \tau_s^4,
\end{eqnarray}
which can numerically be checked to be a saddle point. Therefore we need to
include the $\alpha'$ corrections.

 We know that including the  $\alpha'$
corrections to the scalar potential gives rise to
nonsupersymmetric AdS minima before any lifting allows a de Sitter
lifting.
 One must
then add the lift in a controlled manner to avoid wiping out the minimum
altogether. This amounts to assuming a fine tuning of the coefficient
of the uplifting term with respect to the coefficient of the $\alpha'$
correction term as in the original KKLT scenario. Given that the cosmological
constant of the AdS
minimum is roughly $-1/{\cal{V}}^3$, and that the lifting term is
of the form $1/{\cal{V}}^{8/3}$, it can be seen that a fine tuning of order
$1/{\cal{V}}^{1/3}$ is required.


We checked numerically that lifting to a stable de Sitter vacuum
can indeed be achieved, with values $a_s = 2, \lambda = 4, \mu=20,
\xi=131, \nu = 3.125.$ The resulting volume is roughly $50000$,
with ${\cal{V}}^{-1/3}\approx 0.06$ and $\nu/\xi\approx 0.02.$


To achieve a fine tuning of this magnitude, a hierarchically small FI term is
required. This may be achieved already due to the suppression factor of
$4\pi^2$
in the FI term. It may also be the result of warping: the low energy effective
action for the magnetised D7 brane contains a term coming from the
Yang-Mills kinetic terms, of the form
\begin{equation}
\label{ymterm}
\int d^8 x \sqrt{\det{g_8}} F_{mn} F^{mn},
\end{equation}
where $g_8$ is the induced metric on the D7 brane. Assuming a warped
ansatz for the metric of the form
\begin{equation}
ds^2 = e^{2 A(y)} \eta_{\mu\nu} dx^\mu dx^\nu +
e^{-2 A(y)} g_{mn} (y)dy^m dy^n,
\end{equation}
we observe that the powers of $e^{2 A(y)}$ cancel out of
$\sqrt{\det{g_8}}.$ Assuming that the
D7 brane has a constant warping $A(y) = A$ over its worldvolume,
the raising of indices in $F^{mn}$ gives $e^{4 A}$, indicating that the term
(\ref{ymterm}) is suppressed by an overall warp factor.

This effect may be hard to achieve in practise since the
4-modulus $T_b$ corresponds to the overall volume of the Calabi-Yau,
and it is difficult to envisage a situation in which most of that 4-cycle
can be in a highly warped region.

\subsubsection{Fluxes on the small cycle}

\label{smallcyclesec}
Let us now consider turning on magnetic flux on a brane wrapping one
of the small 4-cycles. We need to include the dependence of the K\"ahler metrics on the small modulus. Following \cite{CCQPaper}, 
we parametrise this dependence as
\begin{equation}
K = -2 \log {{\cal{V}}} + {|\rho|^2\over {(T_s + T_s)^{\beta}
    {\cal{V}}^\alpha}},
\end{equation}
where $T_s$ is the modulus of the small 4-cycle. Let us estimate first the
contribution to the F-term energy, coming from $e^K K^{\rho \rho}
| D_\rho W|^2.$ This can be seen to scale as
\begin{equation}
\label{oldfterm}
| \rho|^2 |W_0|^2 {1\over{{\cal{V}}^{2+\alpha} (T_s+T_s^*)^\beta}}.
\end{equation}
In fact, a careful analysis similar to the one in Appendix \ref{coefcomp}
shows that the F-term contribution is
\begin{equation}
\label{totalfterm}
\left(1-{3\alpha\over 2}+8\beta\right)|\rho|^2 |W_0|^2 {1\over{{\cal{V}}^{2+\alpha}
(T_s+T_s^*)^\beta}}.
\end{equation}
If $\beta$ is negative, and larger than $1/8$ in magnitude, the overall
coefficient
in (\ref{totalfterm}) will be negative.

The superpotential for this case may be written as
\begin{equation}
\label{superpot2}
W = W_0 + \rho \varphi \tilde\varphi + A_s {e^{-a_s T_s}
\over {\left(\det(\varphi  \tilde\varphi)\right)^p}}
\end{equation}
with $p$ the corresponding power in the Affleck-Dine-Seiberg superpotential. 
Note that we do not include a term allowed by gauge invariance,
$\rho^{a} e^{-b T_s}$, since the $\rho$ fields are not charged under the
nonabelian gauge group which condenses and hence are not expected to appear
in the superpotential (note that once we integrate out the $\varphi$'s such coupling will appear).

As mentioned in section \ref{spectrumsec}, there are two possibilities for
which of the K\" ahler
moduli ($T_b$ and $T_s$) becomes charged under the anomalous $U(1).$
Let us first consider the case that $T_s$ becomes charged.
The charges of the $\varphi$ fields relative to the charge of $T_s$ under the
anomalous $U(1)$ are fixed by the gauge invariance of $(\ref{superpot2}).$
The sign of the FI term may or may not change from the case of fluxes on the
large cycle, depending on whether $(\partial_{T_s} K)$ has the same or opposite
sign to
$(\partial_{T_b} K).$

In the case that the $\varphi$ fields have positive mass squared, we
can integrate them out keeping only $\rho.$
Then if $\rho$ has positive mass squared, the lifting can be
achieved since the D-term scales
as $1/{\cal{V}}^2.$ If however, $\rho$ is tachyonic, then after minimising with
respect to $\rho$, the
combination of (\ref{totalfterm}) and the D-term potential gives a negative
energy
contribution, not allowing for a de Sitter minimum. 

In the large $\cal{V}$ limit, the D-term potential is
\begin{equation}
{1\over{T_s+T_s^*}}\left( {r\over{\cal{V}}} - {|\rho|^2
  q\over{{\cal{V}}^\alpha (T_s + T_s^*)^\beta}} \right)^2
\end{equation}
Minimising the sum of F-term and D-term contributions to the energy
with respect to $\rho$ gives
\begin{equation}
\left(1-{3\alpha\over 2}+8\beta\right)\cdot{|W_0|^2\over{{\cal{V}}^{2+\alpha} (T_s + T_s^*)^\beta}}
= {2q \over {{\cal{V}}^\alpha (T_s + T_s^*)^{\beta+1}}} D,
\end{equation}
with
\begin{equation}
D \equiv \left( {r\over{\cal{V}}} - {|\rho|^2
  q\over{{\cal{V}}^\alpha (T_s + T_s^*)^\beta}} \right)
\end{equation}
In the large volume limit one has $D\sim
1/{\cal{V}}^2.$
Therefore, the D-term potential behaves like $D^2 \sim 1/{\cal{V}}^4.$

The VEV of $\rho$ is fixed by
\begin{equation}
| \rho|^2 \sim {r {\cal{V}}^\alpha\over{q \cal{V}}}.
\end{equation}
The F-term contribution to the energy is therefore $\sim
1/{\cal{V}}^3$ and is insufficient to lift the vacuum to de Sitter, even
if its overall coefficient is positive.
Interestingly, this behaves like $\alpha'$ corrections
to the scalar potential and could be used for proving the existence
of large volume minima for Calabi-Yau manifolds $M$ with $\chi (M) =
2(h^{1,1} - h^{1,2}) >0.$ This is a very model dependent statement, though,
since one requires $r/q$ to be large enough.

If the $\varphi$ fields have opposite charge to that of the FI term,
they cannot be integrated out,
and since they can cancel the D-term, a similar argument to the one above
applies
so that lifting cannot be achieved.

The following case is that the field that becomes charged under the anomalous
$U(1)$ is
$T_b$ rather than $T_s$. Anomaly cancellation implies then that
there cannot be $\varphi$, $\ti \varphi$ fields living between magnetised and unmagnetised branes wrapping the small 4-cycle.
The superpotential is thus given by (\ref{superpot}).
Since the sign of the FI term is determined by $\partial_b K$, it is negative. In case there are branes wrapping the large cycle, an anomaly is generated, and consequently 
$\varphi$ fields must exist (this time corresponding to strings stretched between the branes
on the small
4-cycle and those wrapping the large 4-cycle) and will have a positive mass. After
integrating them out,
the D-term potential is of the form
\begin{equation}
{1\over{T_s+T_s^*}}\left( {r\over{{T_b+T_b^*}}} \pm {q|\rho|^2
  \over{{\cal{V}}^\alpha (T_s + T_s^*)^\beta}} \right)^2
\end{equation}
If the sign of $\rho$ is positive, lifting can be achieved with an appropriate
amount of warping. If it is negative, and the sign of the F-term contribution
(determined
by $1-3\alpha/2+8\beta$) is negative, minimising with respect to $\rho$ will only
yield a negative total contribution. Notice that this will likely be the case if $\alpha>0$ and $\beta<0$, as advocated in \cite{CCQPaper}. In this case, lifting would be impossible. 
However, the result of \cite{CCQPaper} relies on the assumption that the matter fields localised in the small cycle decouple from the dynamics of the large cycle, which will
not generically be the case if the two cycles intersect and there are branes wrapping both of them.

The last case we will analyse is having magnetised branes wrapping the large cycle in such a way that it is the small cycle the one that gets charged. This case is rather similar to
the one with fluxes on the small cycle with the small cycle becoming charged, and, similarly to that case, it does not give de Sitter vacua.


\section{Conclusions}
We have studied the spectrum of chiral fields in magnetised D7
branes, with a gauge group $SU(N_c)\times U(1)$ and fields
$\varphi,\tilde{\varphi}$ in the fundamental of $SU(N_c)$ and anomalous $U(1)$
charge $+1$, and $SU(N_c)$ singlets
$\rho$ with anomalous $U(1)$ charge $\pm 2$.
This has allowed us to consider  several  scenarios
depending on the number of moduli and the location of the D7 branes
and magnetic fluxes.

\begin{enumerate}
\item{}
If all matter fields have positive charge, then the D-term cannot vanish
and the minimum of the scalar potential can be lifted to de Sitter
space. This can be considered a string realisation of the model
considered by Ach\'ucarro {\it et al} \cite{beatriz}.

\item{}
The fields $\varphi, \tilde{\varphi}$ have positive charge but the fields
$\rho$
have negative charge. In this case the KKLT-like scenario with
supersymmetric AdS minimum remains essentially unchanged and the D-term does
not
lift the minimum to de Sitter space, in full agreement with the
arguments of Choi {\it et al}. However once $\alpha'$ corrections are
included both F- and D-terms are non-vanishing and the minimum may or may not
be
lifted to de Sitter depending on the type of 4-cycle that the D7
branes wrap.
\end{enumerate}

The lifting mechanism works naturally for volumes of order in the
thousands, but in order to work properly for larger volumes, the 2-cycle in which the
magnetic flux is turned on has to be at the tip of a deeply warped
throat. The lifting is achieved by tuning the warp factor
appropriately as in the original KKLT scenario.
The necessary geometry is possible to realise
\footnote{See for
  instance \cite{ouyang}. We thank H. Verlinde for discussions on this point. }
but an explicit construction is beyond the scope of this article.

In the cases where the D-term lifting does not happen, a relevant question
would be if the D-terms change the nature of the minima found in the
absence of magnetic fluxes. We have seen that, as expected, the
supersymmetric
AdS minima such as the original KKLT scenario are not affected by the
anomalous $U(1)$. Furthermore
the same happens for large volume minima. The fact that there are
minima with exponentially large volume remains true even after adding
the D-terms, independent of the fact that they lift to de Sitter space
or not. This makes these scenarios more robust.

Our mechanism of moduli stabilisation can be seen as a generalisation
of the mechanism proposed in \cite{mp} of using only D-terms and soft
supersymmetry breaking terms to stabilise the K\"ahler moduli. In
principle we could achieve this by turning-off the non-perturbative
effects\fn{A related mechanism of Kahler moduli stabilisation with magnetic fluxes in toroidal setups has been proposed in \cite{tristan}.}.
However when we turn them off, the field equations for the K\"ahler
moduli become linear and they are not stabilised. Introduction of (anti)
D9-branes may be needed to achieve stabilisation but then care most be taken
about avoiding Freed-Witten anomalies. This is clearly model dependent
and will not be addressed further here.

We would like to point out that a crucial part of our calculations was
to use the recently computed K\"ahler moduli dependence of matter
fields
\cite{CCQPaper}. It was crucial to know the modular weights of the
matter fields in order to establish the positivity of the contribution
of the matter fields to the F-term part of the potential. Here we used
the simplest case in which all matter fields are assumed to wrap the
same cycle. Other cases discussed in \cite{CCQPaper} are easily
incorporated.

In most of our considerations we only included a single matter field
$\rho$ singlet under the non-abelian gauge symmetry. In general it is
a matrix
 of these fields $\rho_{ij}$ with $i,j$ flavour indices.
In the nonperturbative superpotential, the term $\rho^{N_F}$ means
actually $\det \rho_{ij}$ and in the D-terms and K\"ahler potential,
these fields appear in the combination $\sum_{ij} |\rho_{ij}|^2$. This
means that there are several combinations of these fields that do not
appear explicitly in the potential and could remain flat. It would be
interesting to study their possible role as inflaton candidates once
further corrections to the potential are included. In any case there
is no problem about their stabilisation since they are bounded complex
quantities which are always stabilised at finite values.

In summary, we have generalised the current discussions on moduli
stabilisation to include magnetised D7 branes of the type expected to
include the standard model in a fully realistic setting. We hope this
is only a first step towards a more stringy realisation of realistic
chiral models within the KKLT and exponentially large volume scenarios. The
fact
that the D-terms can be actually used for de Sitter lifting is very encouraging.
Open
questions regarding the actual structure of soft supersymmetry
breaking terms remain to be discussed in detail. In particular, the
F-term of the matter field $\rho$ is generally non-zero and could
contribute to the structure of soft supersymmetry breaking in the
observable sector.
Here we can only say
that, as already discussed in the literature, the lifting mechanism
is the leading source of supersymmetry breaking in KKLT models
\cite{hepth0503216,choi} but its
effect on the exponentially large volumes is less relevant  due to the fact
that the original AdS minimum is already non-supersymmetric
\cite{hepth0505076,largev}.

With this work we believe to have clarified a number of important
issues regarding the effects of anomalous $U(1)$'s on the moduli
stabilisation procedure. This puts  the mechanisms of
moduli stabilisation on firmer grounds and also allows the possibility for de
Sitter
lifting in a controllable manner. The explicit realisation of the
lifting mechanism is model dependent, which we have illustrated with
some representative cases. We hope these techniques and results will be useful
in detailed constructions of realistic string models including moduli
stabilisation.

\section{Acknowledgements}
We thank S. Abdussalam, C.P. Burgess, B. de Carlos, A. Casas, J. Conlon, G.
Dvali,
S. Giddings, L.E. Ib\'a\~nez, F. Marchesano, G. Tasinato,
 A.Uranga, H. Verlinde,
I. Zavala for useful discussions on the subject of this
 article.
DC, MPGM and FQ thank the KITP Santa Barbara and organizers of the
`String Phenomenology' workshop where part of this research was done.
MPGM thanks Perimeter Institute and
CECS, Valdivia, Chile, where part of this work was done,
for kind hospitality and support. The work of DC is supported by the University of Cambridge.
MPGM is partially supported by the
European Community's Human Potential Programme under contract
MRTN-CT-2004-005104 and by the Italian MUR under contracts PRIN-2005023102
and PRIN-2005024045. FQ is partially funded by PPARC and a Royal Society
Wolfson merit
award. KS is grateful to Trinity College, Cambridge, for financial support.

\appendix

\section{Anomaly cancellation}

We check in this section\fn{This section follows closely refs. \cite{csu,fdo_thesis}. Though these references dealt only with toroidal and orbifold constructions, their results can be extended to more general constructions since they rely only in the bi-linearity of the intersection product between D7 and D5 charges.}
how the matter content described in section
\ref{spectrumsec}, together with the GS mechanism, cancels all possible
anomalies that could be present in our setup.

\subsection{$U(1)_a-SU(N_b)^2$ anomalies}
The mixed anomaly is given by the formula
\beqa
\cA_{ab}=\sum_r Q_a(r) \cdot C_b(r)
\eeqa
where $C_b(r)$ is the quadratic Casimir of $SU(N_b)$, and $Q_a(r)$ is the
$U(1)$ charge.
Let us consider first the case $a\neq b$. The only fields
relevant for this computation are the $\varphi$ and $\ti \varphi$. Knowing that
for both fundamental and anti-fundamental representations one has $C(r)=\oh$,
one readily gets
\beqa
\cA_{ab}=\oh N_a(I_{ab}+I_{ab'}).
\label{remainder1}
\eeqa
This anomaly should eventually be cancelled by a GS term. Now, consider the
case $a=b$. Now, the relevant fields are not
only the $\varphi$'s and $\ti \varphi$'s but also the symmetrics and the
antisymmetrics. Knowing that
\beqa
C(\Ysymm_{a})&=& {N_a+2\over 2},\\
C(\Yasymm_{a})&=& {N_a-2\over 2},\\
Q(\Ysymm_{a})&=&Q(\Yasymm_{N_a})=+2,
\eeqa
we get
\beqa
\cA_{aa}&=&\oh \sum_{c\neq a}N_c (I_{ac}+I_{ac'})+2\cdot{N_a+2\over 2} (\#
\Ysymm)+2\cdot{N_a-2\over 2} (\# \Yasymm)\nonumber\\
&=&{1\over 2b} \left[bN_aI_{aa'}+b\sum_{c\neq
a}N_c(I_{ac}+I_{ac'})-I_{aO}\right] + \oh  N_a I_{aa'}.
\eeqa
Now, the term between square brackets vanishes because of (\ref{tadpoles}), and
the remaining anomaly is just $\oh N_a I_{aa'}$. It has the same form as
(\ref{remainder1}), so we can say in full generality that the leftover anomaly
$U(1)_a-SU(N_b)$ anomaly between some $a$ and $b$ sectors is just given by
\beqa
\cA_{ab}=\oh N_a(I_{ab}+I_{ab'}).
\label{mixed_an}
\eeqa
This is the term that must be cancelled by the GS mechanism\fn{We recommend
the reference \cite{preskill} for a nice and detailed explanation of how the
GS mechanism takes place from the field theory point of view.}. It is easy to
see that it is indeed the case. As explained in the main text, magnetic flux in
the world-volume of the
D-brane $a$ generates a charge for every modulus $T_i$ which intersects the
two-cycle(s) where magnetic field has been turned on. Wrapping a D7-brane in
the four cycle
associated to $T_i$ induces an anomaly for the $U(1)_a$. The anomaly is given
precisely by the products of the coefficients of $\int_{M^4} D_{2,i}\wedge F_a$
and
$\int_{M^4} \pim T_i F_i\wedge F_i$, with $D_{2,i}$ being the Hodge dual in
four dimensions of $\pim T_i$, and $F_i$ the $SU(N_i)$ field strengths.
But this coefficient is precisely of the form (\ref{mixed_an}), since it is
equal to the D5 charge of the brane $a$ times the D7 charge of the brane $i$.
Note that the factor
$N_a$ arises in the CS term from the normalisation of the $U(1)$.

\subsection{$U(1)_a-U(1)_b^2$ anomalies}
Let us analyse the case of the cubic anomaly. Let us start by assuming $a\neq
b$. This anomaly is given by
\beqa
\cA_{ab}=\sum_r Q_a(r)\cdot Q_b^2(r).
\eeqa
In this case only the $\varphi$ and $\ti \varphi$ fields are relevant and one
gets
\beqa
\cA_{ab}=N_aN_b (I_{ab}+I_{ab'})
\label{anom2}
\eeqa
The factor $N_aN_b$ comes since in a bifundamental representation there are
$N_aN_b$ fields from the $U(1)$ point of view. In the case $a=b$, we have
to take into account that there a number of $\rho$ fields given by (\ref{nrho})
with charge $+2$ (again, in case the number is negative, it means that the
charge is $-2$ so
that everything is consistent). In this case the anomaly reads (the 1/3 factor
comes from the symmetry of the diagram)
\beqa
\cA_{ab}&=&{1\over 3}\left(N_a\sum_{c\neq a}N_c (I_{ac}+I_{ac'})+2^3\left[{
N_a^2\over 2} I_{aa'}-{N_a\over 8b}I_{aO}\right]\right)\nonumber\\
&=&{N_a\over 3b}\left[bN_aI_{aa'}+b\sum_{c\neq
a}N_c(I_{ac}+I_{ac'})-I_{aO}\right]+ N_a^2 I_{aa'}.
\eeqa
Again, the term in square brackets vanish because of (\ref{tadpoles}) and the
remainder is $N_a^2 I_{aa'}$, whose form is the same as that of (\ref{anom2}).
We can
write in full generality that the remainder of the cubic anomaly
$U(1)_a-U(1)^2_b$ is given by
\beqa
\cA_{ab}=N_aN_b(I_{ab}+I_{ab'}).
\eeqa
Again, this term will be cancelled by the GS mechanism. It is easy to see that
this is the case, following a similar argument to that sketched for the mixed
anomaly case.
Note that in this case there is an extra factor of $N_b$ appearing from the
cubic case due to the fact that we are considering the coupling to the $U(1)_b$
field, coming from a term
$\sim N_b \int_{M^4} F_b \wedge F_b$, $F_b$ being in this case the $U(1)$ field
strengths.

\subsection{Gravitational anomalies}
The gravitational anomaly for a given $U(1)_a$ group is just given by
\beqa
\cA_a=\sum_r Q_a(r) = \Tr Q_a.
\eeqa
In our case this is given by
\beqa
\cA_a=N_a\sum_{c\neq a}N_c(I_{ac}+I_{ac'})+2\left({N_a^2\over
2}I_{aa'}-{N_a\over 8b}I_{aO}\right).
\eeqa
After imposing the tadpole cancellation we obtain
\beqa
\cA_a={3\over 4b}N_aI_{aO}.
\eeqa
This is the bit that has to be cancelled by the GS mechanism. One can check it does, following the steps outlined in \cite{csu}.

\section{Metrics on moduli space}
\label{metricsappendix}

We now compute the metric and inverse metric on moduli space, in the
limit of (the real part of) $T_b$ large. The relationship between the
volume and $T_b$ is (for large $T_b$, and up to constant factors)
${\cal{V}} \sim T_b^{3/2}.$
\begin{equation}
\label{metricwithrho}
K_{i\bar{\jmath}} =
\left( \begin{array}{ccc}
1/{\cal{V}}^{4/3} & 1/{\cal{V}}^{5/3} & 1/{\cal{V}}^{{2\over3}(\alpha + 1)} \\
1/{\cal{V}}^{5/3} & 1/{\cal{V}} & 0 \\
1/{\cal{V}}^{{2\over3}(\alpha+1)} & 0 & 1/{\cal{V}}^{{2\over3}\alpha}
\end{array} \right).
\end{equation}
Here the moduli are put in order $\{ 1,2,3 \} = \{ T_b, T_s, \rho \}.$

The determinant of $K_{i\bar{\jmath}}$ is
${\cal{V}}^{-{2\over3}\alpha - {7\over3}}.$ The inverse metric is thus
\begin{equation}
K^{i\bar{\jmath}} = {\cal{V}}^{{2\over3}\alpha + {7\over3}}
\left( \begin{array}{ccc}
{\cal{V}}^{ -{{2/3}\alpha - 1}} & {\cal{V}}^{ { - {2/3}\alpha - {5\over3}}}
 & {\cal{V}}^{ { -{2/3}\alpha - {5\over3}}}\\
{\cal{V}}^{- { {2/3}\alpha - {5\over3}}} &
{\cal{V}}^{- { {2/3}\alpha - {4\over3}}} &
{\cal{V}}^{- { {2/3}\alpha - {7\over3}}}\\
{\cal{V}}^{- { {2/3}\alpha - {5\over3}}} &
{\cal{V}}^{- { {2/3}\alpha - {7\over3}}} &
{\cal{V}}^{- {7\over3}}
\end{array} \right).
\end{equation}
So that
\begin{equation}
K^{i\bar{\jmath}} = \left( \begin{array}{ccc}
{\cal{V}}^{4/3} & {\cal{V}}^{2/3} & {\cal{V}}^{2/3} \\
 {\cal{V}}^{2/3} & {\cal{V}} & 1 \\
 {\cal{V}}^{2/3} & 1 & {\cal{V}}^{{2\over3} \alpha}
\end{array} \right).
\end{equation}

Following \cite{hepth0502058}, let us now consider the limit of large
$\cal{V}$ such that ${\cal{V}}\sim e^{a_s \tau_s}.$
The first thing to consider is where the axion of $T_s$ is fixed.
The relevant terms, before including the effects of $\rho$,  are the
mixed terms from $K^{s\bar{b}} D_s W \overline{D_b W}$, of the form
$K^{s\bar{b}} \overline { (\partial_b K) W} (\partial_s W).$
This term scales as
$ W (\partial_s W)$ with $\cal{V}.$ After including the terms with $\rho$,
there are more contributions to the axion potential, coming from
the mixed terms in $K^{s\bar{\rho}} (D_s W) \overline{(D_\rho W)}$,
$K^{b\bar{\rho}} (D_b W) \overline{(D_\rho W)}$,
$K^{\rho\bar{\rho}} (D_\rho W) \overline{(D_\rho W)}$, and their
complex conjugates.
The parts of these terms involving $\partial_\rho W$ are all vanishing
as we are assuming there is no nonperturbative contribution involving
$T_b, \rho$ in the superpotential.
Therefore, from the first term, only
$K^{s\bar{\rho}} (\partial_s W) \overline{(\partial_\rho K) W}$
is relevant, scaling as
${\cal{V}}^{-{2\over3} \alpha} (\partial_s W) W.$

Thus the axion is still fixed mainly by the
$K^{s\bar{b}} \overline{(\partial_b K) W} (\partial_s W)$ term,
at a value which makes the overall sign of this term negative.

\section{Computing the $|\rho|^2$ F-term contribution}
\label{coefcomp}
The determinant of the metric $K_{i\bar{\jmath}}$ in (\ref{metricwithrho}) can
be
seen to scale as $K_{bb} K_{ss} K_{\rho \rho} + K_{\rho b} K_{b\rho} K_{ss}.$
These terms can be estimated individually as
\begin{eqnarray}
K_{b\bar{b}} &=& {3\over{\cal{V}}^{4/3}} + {{\alpha (\alpha+1) |\rho|^2}\over
{{\cal{V}}^{2(\alpha+2)/3}}} \\
K_{s\bar{s}} &=& {3\over{2{\cal{V}} (T_s+T_s^*)^{1/2}}}\\
K_{\rho \bar{\rho}} &=& {1\over{{\cal{V}}^{2\alpha/3}}}\\
K_{b\rho} &=& -{\alpha\rho\over{\cal{V}}^{2(\alpha+1)/3}}.
\end{eqnarray}

The first contribution which includes a $|\rho|^2$ factor comes from the term
in the scalar potential
\begin{equation}
K^{\rho\bar{\rho}} |(\partial_\rho K) W|^2,
\end{equation}
and the contribution turns out to be
\begin{equation}
{|\rho|^2\over{\cal{V}}^{2\alpha/3}} |W|^2.
\end{equation}
The next contribution is from $K^{\rho \bar{b}} (\partial_\rho K) W
\overline{(\partial_b K) W}$, and turns out to give
\begin{equation}
- {{\alpha |\rho|^2}\over{{\cal{V}}^{2\alpha/3}}}.
\end{equation}
There is an identical contribution from the conjugate of this term.

The next contribution comes from $K^{\rho \bar{s}} (\partial_\rho K) W
\overline{(\partial_s W) W}$, whose scaling with volume is
\begin{equation}
-{{3\alpha (T_s+T^*_s)^{3/2}|W|^2 |\rho|^2\over{{\cal{V}}^{2\alpha/3 +
13/3}}}}.
\end{equation}

The following term to consider is $K^{b\bar{b}} (\partial_b K) W
\overline{(\partial_b K) W}$ (there is a term in the inverse metric
appearing in $K^{bb}$ containing $|\rho|^2$, and also two more terms
coming from the $(\partial_b K)$ factors). The scaling is given by
\begin{equation}
{\alpha\over{\cal{V}}^{2\alpha/3}} |\rho|^2 |W_0|^2
\end{equation}

The next term is $K^{s\bar{s}} (\partial_s K) W
\overline{(\partial_s K) W}$, giving
only terms which scale as
\begin{equation}
{1\over{{\cal{V}}^{2\alpha/3+1/3}}},
\end{equation}
which is suppressed with respect to the contributions we computed before.

Lastly, one may also consider $K^{s\bar{b}} (\partial_s K) W
\overline{(\partial_b K) W}.$ This gives
\begin{equation}
{|\rho|^2\over{{\cal{V}}^{2\alpha/3 + 2/3}}}.
\end{equation}
In total, after taking into account the pre-factor of $e^K\sim 1/{\cal{V}}^2$,
the highest order F-term contribution including a $|\rho|^2$ factor is
\begin{equation}
(1-\alpha){|\rho|^2|W_0|^2\over{{\cal{V}}^{2\alpha/3}}}.
\end{equation}
This is still positive for $\alpha=1/2.$

\section{FI terms in the toroidal case}
\label{app_torus}
The main aim of this appendix is to fix the sign of the FI term with respect to
the charge of the $\varphi$ fields. To do this, we will compute
the mass of the $\varphi$'s from the purely stringy formulae and check they
consistently come from a FI term. Then we check that the corresponding
FI term computed by other means agrees with this calculation. The content of this section is based on ref. \cite{qsusy}.

\subsection{From the mass formula}

Consider a factorisable $T^6=(T^2)^3$ whose basis of 1-forms we denote by
$dx^i$, $dy^i$, $i=1,2,3$.
A IIA factorisable $D6$ brane in such toroidal setup is described by six
wrapping numbers\fn{See \cite{fdo_thesis} for a review on intersecting branes.}
~\\
\begin{center}
\begin{tabular}{cccc}
$D6$&$(n^1,m^1)$&$(n^2,m^2)$&$(n^3,m^3),$
\end{tabular}
\end{center}
~\\
where $n_i$ ($m_i$) represents the number of times the brane wraps the $x^i$
($y^i$) direction. Given two stacks of D6 branes that intersect at several points in the $T^6$, there will be a single massless chiral fermion living in each one of the intersection points. The net number of intersection points gives thus the net number of chiral fermions, that is given by the absolute value of
\beqa
I_{ab}=\prod_{i=1}^3(n_a^im_b^i-n_b^im_a^i).
\eeqa
For each one of these intersection points there will be a tower of scalars, of which the lightest ones have masses
\beqa
m_{1,ab}^2&=&{1\over
2\a'}(-|\theta_{ab}^1|+|\theta_{ab}^2|+|\theta_{ab}^3|)\nonumber\\
m_{2,ab}^2&=&{1\over 2\a'}(+|\theta_{ab}^1|-|\theta_{ab}^2|+|\theta_{ab}^3|)\\
m_{3,ab}^2&=&{1\over
2\a'}(+|\theta_{ab}^1|+|\theta_{ab}^2|-|\theta_{ab}^3|)\nonumber\\
m_{4,ab}^2&=&{1\over \a'} \left[1-\oh (+|\theta_{ab}^1|+|\theta_{ab}^2|+|\theta_{ab}^3|)\right],
\eeqa
where $\theta_{ab}^i$ is the angle\fn{Given the wrapping numbers $(n^j_i,m^j_i)$ with all the $n^j_i\geq 0$, this angle is given by $\pi\theta_{ab}^i=\tan^{-1}\left({m^b_i R^y_i\over n^b_i R^x_i}\right)-\tan^{-1}\left({m^a_i R^y_i\over n^a_i R^x_i}\right)$. This is the situation usually stressed in the literature. However, one has to take into account that when negative values for one or both of the $n^j_i$ are used the formula for this angle is slightly different, though straightforward to obtain in a case-by-case analysis.} the brane $a$ makes with the brane $b$, in units of $\pi$.

For arbitrary wrapping numbers, a D6 will go to a (stack of) D9(s)
with magnetic flux under 3 T-dualities along the $x$ directions, but special
choices of wrapping numbers
can lead us to lower dimensional (stacks of) branes. We can use this D6-brane
language to
describe D7 branes, as follows. A $D7_i$ is defined as a D7 that is pointlike
in the $i^{th}$ torus and wraps completely the $j^{th}$ and $k^{th}$ ones.
After the three T-dualities we can describe a single $D7_i$ by the numbers
~\\
\begin{center}
\begin{tabular}{cccc}
$D7_1$&$(1,0)$&$(n^1_2,1)$&$(n^1_3,-1)$\\
$D7_2$&$(n^2_1,-1)$&$(1,0)$&$(n^2_3,1)$\\
$D7_3$&$(n^3_1,1)$&$(n^3_2,-1)$&$(1,0)$
\end{tabular}
\end{center}
~\\
Consider the intersection between a $D7_i$ and a $D7_j$ where we
only turn on magnetic flux in the torus where both D7 branes overlap
(that is, $n_i^j=n_j^i=0$ for these particular $i$, $j$).
Then it is easy to see that there are two lightest scalars, whose
mass is \beqa m_{ij}^2={1\over 2\pi\a'}\left| \arctan
\left({(2\pi)^2\a'\over A_k}n_k^i\right)+\arctan
\left({(2\pi)^2\a'\over A_k}n_k^j\right)\right|\simeq
{2\pi|n_k^i+n_k^j|\over  A_k}. \label{mass_} \eeqa This mass can
never be tachyonic. On the other hand it is clear that since
whenever $n_k^i+n_k^j\neq 0$ we will have a chiral fermion living
between both branes, the presence of this relative magnetic flux is
breaking supersymmetry. We can see that the mass of this scalar can
be seen as coming from a couple of FI terms, each one of them
associated to one of the stacks of branes and, in particular, to the
presence of magnetic flux on them. The FI term associated to a
$D7_i$ brane when magnetic flux is present in the $k^{th}$ torus is
easily computable and given by \beqa \xi_i={2\pi n_k^i\over
A_k}={\int_{T^2_k}F\over A_k}. \label{FI_mass} \eeqa Since the
squared mass (\ref{mass_}) is always positive, one can see that the
sign of the FI term is correlated with the charge of the
bifundamental scalars under the gauge groups. A straightforward
analysis shows that whenever one has two $D7$s that overlap over a
$T^2$ where both of them have flux, the charge under $U(1)_i$ of the
state going from the brane $i$ to the brane $j$ has the same sign as
the FI term associated with the brane $i$.

Now we provide more evidence that this is indeed a FI term, following methods
already developed in papers \cite{qsusy,hepth0609211}, so we will be brief.

\subsection{From the DBI action}
Another way of checking that magnetic flux gives rise to a FI term is to
consider the difference between the DBI action for the D7-brane with magnetic
flux and this same action
in the absence of flux. The DBI action in the Einstein frame reads (notice the
absence of curvature contributions in the toroidal case)
\beqa
S=-T_p\int d^{p+1}x\sqrt{g_{\mu\nu}+2\pi\a' F_{\mu\nu}},
\eeqa
with
\beqa
T_p={(2\pi)^{-p}\a'^{-(p+1)/2}\over g_s}.
\eeqa
From this expression we read the gauge coupling constant (in the supersymmetric
limit) for the $U(1)$ living in the D7:
\beqa
{1\over g_i^2}={V_4\over (2\pi)^5g_s\a'^2}
\eeqa
with $V_4$ the (dimensionful) volume wrapped by the D7. Now, note that the
difference in vacuum energy in the four dimensional theory due to the presence
of the
magnetic flux is
\beqa
\delta V={(2\pi)^{-7}V_4\over g_s\a'^4}\left(\sqrt{1+(2\pi\a')^2\left({2\pi
n_k^i\over A_k}\right)^2}-1\right)\simeq \oh {1\over g^2}
\left({\int_{T^2_k}F\over A_k}\right)^2
\label{FI_DBI}
\eeqa
in the dilute flux approximation. Note that equations (\ref{FI_mass}) and
(\ref{FI_DBI}) can consistently come from a D-term potential of the form
\beqa
V_{D,i}={g_i^2\over 2} \left({\xi_i\over g_i^2}+\sum_j q_j |\phi_j|^2\right)^2
\eeqa
where $\phi_j$ stands for all the canonically normalised fields charged under
the corresponding $U(1)$ with $q_j$ being their corresponding charges, and
$\xi_i$ is
given by (\ref{FI_mass}).

\subsection{From the $N=1$ supergravity algebra}

FI terms can be extracted from the K\"ahler potential $K$ of the
compactification, via the supergravity formula
\beqa
{\xi_i\over g_i^2}=\left({\p K\over \p V^i}\right)_{V^i=0}
\label{formula}
\eeqa
where $V^i$ is the vector superfield of the corresponding $U(1)$. In toroidal
compactifications the K\"ahler potential for closed string moduli reads (both in
Type IIA and
Type IIB)\fn{Note a different normalisation of the K\"ahler potential with
respect to e.g. \cite{qsusy}. In particular \beq
(M_P^2)_{[\rm ours]}={1\over 8\pi}(M_P^2)_{\cite{qsusy}.}
\eeq}.
\beqa
K/M_P^2=-\log(S+S^*)-\sum_{i=1}^3\log(T_i+T_i^{*}) -
\sum_{i=1}^3\log(U_i+U_i^{*}),
\label{kahlerpot}
\eeqa
where (in Type IIB)
\beqa
\preal S&=&{1\over 2\pi g_s}\nonumber\\
\preal T_i&=&{A_jA_k\over (2\pi)^5g_s\a'^2}
\eeqa
The exact formula for the $U$ fields will be irrelevant in what follows.
Now, suppose a $D7_i$ brane wraps the four-cycle whose volume is parametrised
by $T_i$. Suppose we turn on worldvolume magnetic field in the $j^{th}$
2-torus,
$j\neq i$. Then, following the arguments developed in \cite{hepth0609211} for the
general case, the $T$ fields that will become charged will be the ones
corresponding to
the 4-cycles that intersect the 4-cycle wrapped by the $D7_i$ in a 2-cycle
threaded by magnetic flux. In this case, it is easy to see that, since the
torus has no
self-intersections, if a $D7_i$ has magnetic flux in the $j^{th}$ 2-torus, then
only the 4-cycle whose volume is parametrised by $T_k$ will become charged
$(i\neq j \neq k)$. That implies that one must modify the K\"ahler potential
(\ref{kahlerpot}) making the change $T_k\to T_k -Q_k V_i$, where $Q_k$ is the
charge
associated to this cycle, so that it remains a gauge invariant function of the
superfields \cite{dsw}. Applying formula (\ref{formula}) we obtain
\beqa
{\xi_i\over g_i^2}=M_P^2{Q_k\over 2~\preal T_k}.
\label{FI_sugra}
\eeqa
Using the fact that $1/g_i^2=\preal T_i$ and that
\beqa
{M_P^2\over 2}={(2\pi)^{-7}V_6\over g_s^2 \a'^4}
\eeqa
with $V_6$ being the dimensionful compactification volume, we obtain, equalling
(\ref{FI_mass}) and (\ref{FI_sugra})
\beqa
Q_k={(2\pi)^7g_s^2\a'^4\over V_6}\preal T_k \preal T_i {\int_{T_j^2}  F\over
A_j}={1\over (2\pi)^3}\int_{T_j^2} F,
\eeqa
so the D-term potential reads
\beqa
V_{D,i}={1\over 2~\preal T_i} \left({1\over 4\pi^2}{n^i_j\over 2~\preal
T_k}+\sum_a q_a |\phi_a|^2\right)^2
\eeqa
in accordance with \cite{hepth0609211}.

\end{document}